\documentclass[a4paper]{aa}
\usepackage{graphicx}

\def \rsun {\ifmmode$R$_{\odot}\else R$_{\odot}$\fi}
\def \nh {N${\rm _H}$}
\def \hcm {\hbox {\ifmmode $ atom cm$^{-2}\else atom cm$^{-2}$\fi}}
\def \src {Her\,X-1}
\def\approxgt{\mathrel{\hbox{\rlap{\lower.55ex \hbox {$\sim$}}
        \kern-.3em \raise.4ex \hbox{$>$}}}}
\def\approxlt{\mathrel{\hbox{\rlap{\lower.55ex \hbox {$\sim$}}
        \kern-.3em \raise.4ex \hbox{$<$}}}}

\newcommand {\chisq} {$\chi ^{2}$}

\begin{document}

\title{X-ray observations during a Her X-1 anomalous low-state}

\author{A.N. Parmar\inst{1} \and T.~Oosterbroek\inst{1} \and D. Dal
Fiume\inst{2} \and M. Orlandini\inst{2} \and A. Santangelo\inst{3}
\and A. Segreto\inst{3} \and S.~Del Sordo\inst{3}}

\offprints{A.N. Parmar (aparmar@astro.estec. esa.nl)}

\institute{Astrophysics Division, Space Science Department of ESA, 
ESTEC, P.O. Box 299, 2200 AG Noordwijk, The Netherlands
\and Istituto TESRE, CNR, via Gobetti 101, I-40129 Bologna, Italy
\and IFCAI, CNR, via La Malfa 153, I-90146 Palermo, Italy}

\thesaurus{(02.01.2; 08.09.2; 08.14.1; 13.25.5)}

\date{Submission date: August 1999; Received date: Accepted date:}

\authorrunning{A.N. Parmar et al.}

\maketitle 

\markboth{Anomalous low-state of Her X-1}
{Anomalous low-state of Her X-1}

\begin{abstract}

Results of a 1999 July 8--10 BeppoSAX observation 
during an anomalous low-state of \src\ are presented.
The standard on-state power-law and blackbody continuum model 
is excluded at high confidence unless partial covering is included.
This gives a power-law photon index of $0.63 \pm 0.02$ and
implies that $0.28 \pm 0.03$ of the flux undergoes additional
absorption of $(27 \pm 7) \times 10^{22}$~atom~cm$^{-2}$.
11\% of the observed 0.1--10~keV flux
is from the $0.068 \pm 0.015$~keV blackbody.
1.237747(2)~s pulses with a semi-amplitude of $2.1 \pm 0.8$\%
are detected at $>$99.5\% confidence and confirmed by RXTE measurements. 
This implies that \src\ underwent substantial spin-down
close to the start of the anomalous low-state. 
The spectral and temporal changes are similar to those recently
reported from 4U\,1626-67. These may result from a strongly warped
disk that produces a spin-down torque. The X-ray source
is then mostly viewed through the inner regions of the accretion disk.
A similar mechanism could be responsible for the \src\ anomalous low-states.
Shadowing by such an unusually warped disk could produce observable 
effects in the optical and UV 
emission from the companion star.

\end{abstract}

\keywords{accretion, accretion disks -- Stars: individual: \src\
-- Stars: neutron -- X-rays: stars}  

\section{Introduction}

\src\ is an eclipsing X-ray pulsar with a pulse period, ${\rm P_s} =
1.24$~s and an orbital period of 1.70~days. The source usually exhibits a
35~day X-ray intensity cycle consisting of a $\sim$10 day duration main
on-state and a fainter $\sim$5 day duration secondary, or short, on-state.
At other 35-day phases
\src\ is still visible as a low-level X-ray source (Jones \& Forman
\cite{j:76}). The effects of X-ray heating on the companion are
evident throughout the 35-day cycle (Boynton et al. \cite{b:73}).
The 35-day cycle probably results from a warped
accretion disk that periodically obscures 
the line of sight to the
neutron star while partially shadowing the companion
(Gerend \& Boynton \cite{g:76}).

The 35-day cycle has been evident in RXTE All-Sky Monitor
(ASM) 1.5--12~keV data with the main-on state being clearly
detected every 35 days for $>$3~years (e.g., Scott \& Leahy \cite{s:99}).  
An exception to this
regularity occurred when
the on-state expected around 1999 March~23 was not
detected (Levine \& Corbet \cite{l:99}). Similar exceptions have
been detected twice before.
In 1983 June to August EXOSAT failed to detect an X-ray on-state from \src\
and instead a faint source, with a strength comparable to that of the
low-state emission was observed over a wide range of 35-day phases 
(Parmar et al. \cite{p:85}). Optical observations
during this interval, and in 1999 April, 
detected the effects of strong X-ray heating
on the companion star (HZ~Her)
indicating that it was still being irradiated by a strong X-ray source
(Delgado et al. \cite{d:83}; Margon et al. \cite{m:99}). 
The anomalous low-state lasted $<$0.8~year
with the 35-day cycle returning by 1984 March. 
Similarly, in 1993 August
ASCA failed to observe the expected on-state, again detecting instead a 
faint X-ray source (Vrtilek et al. \cite{v:94}; 
Mihara \& Soong \cite{m:94}). 
We report here on Target of Opportunity
observations of the third known anomalous low-state of
\src.

\section{Observations}
\label{sec:observations}

Results from the Low-Energy Concentrator Spectrometer (LECS), 
the Medium-Energy Concentrator Spectrometer (MECS),
the High Pressure Gas Scintillation Proportional Counter
(HPGSPC) and the Phoswich Detection System (PDS)
on-board BeppoSAX (Boella et al. \cite{b:97}) are presented. 
All these instruments are coaligned 
and referred to as the Narrow Field Instruments, or NFI.

The region of sky containing \src\ was observed by BeppoSAX
on 1999 July 08 08:16 UT to July 10 06:20 UT.
This is just after the expected time of
turn-on to the main-on state three 35-day cycles after the non-detection
by Levine \& Corbet (\cite{l:99}).
Data processing was performed in the standard manner using
the SAXDAS 2.0.0 analysis package.
LECS and MECS data were extracted centered on the position of \src\ 
using radii of 8\arcmin\ and 4\arcmin, respectively.
The exposures 
in the LECS, MECS, HPGSPC, and PDS instruments are 45.5~ks, 58.9~ks,
57.2~ks, and 28.5~ks, respectively. 
Background subtraction for the imaging instruments
was performed using standard files.
Background subtraction for the HPGSPC was 
carried out using data obtained when the instrument
was looking at the dark Earth and for the PDS using data
obtained during intervals when the collimator was offset from the 
source. 

\begin{figure}
\begin{center}
\hbox{\hspace{-0.0cm} \includegraphics[width=8.0cm,angle=0]{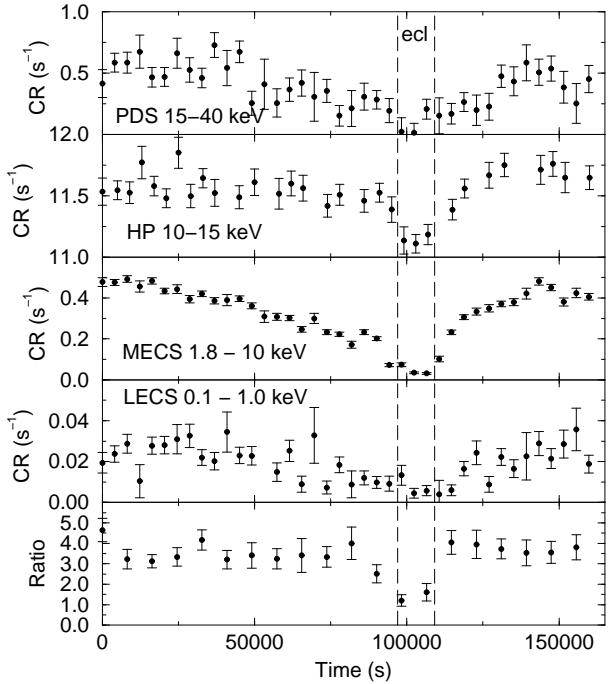}}
\caption[]{Background subtracted lightcurves in 4 energy bands with a binning
         of 4096~s and LECS hardness ratios (4.0--10~keV/0.1--1.0~keV) with a
         binning of 8192~s. CR is count rate. The predicted on-state eclipse
         ingress and egress are indicated with dashed lines. Time is
         from the observation start}
\label{fig:lc}
\end{center}
\end{figure}

\section{Analysis and Results}
\label{sec:analysis}

\subsection{X-ray Lightcurve}
\label{subsec:lc}

Background subtracted lightcurves in 4 energy ranges
and the hardness ratio
(LECS counts between 4.0--10~keV divided by those between
0.1--1.0~keV) are shown in Fig.~\ref{fig:lc}. 
The eclipse ingress and egress, based on the ephemeris
of Deeter et al. (\cite{d:91}), and assuming an on-state eclipse duration
of 5.5~hr, is indicated. The lightcurves show a gradual reduction 
in flux before the eclipse and a more rapid increase following
egress. The eclipse is not total with significant flux remaining. 
There is no evidence for any significant change in
hardness ratio, expect when the eclipse and non-eclipse intervals 
are compared.

\begin{figure}
\includegraphics[height=8.0cm,angle=-90]{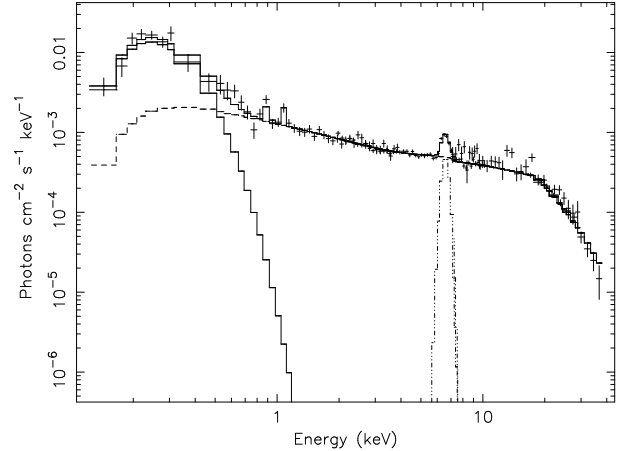}
\caption[]{The NFI \src\ non-eclipse spectrum.
The solid line shows the unfolded spectrum 
obtained with a partially covering absorber (see
Table~\ref{tab:spectral_fits}). 
The contributions from the blackbody, power-law,
and Fe K line are indicated separately}
\label{fig:spectra}
\end{figure}

A period search in the range 1.23770--1.23776~s,
predicted by extrapolating the spin-period history 
in Bildsten et al.\ (\cite{bi:97}),
was performed using the combined LECS and MECS data.
The photon arrival times were converted to the solar barycenter
and additionally to the center of \src\ mass using the ephemeris of
Deeter et al.\ (\cite{d:91}).  
A peak at ${\rm P_s}$=1.237747~s was found
with a \chisq\ of 42 for 16 degrees of
freedom (dof). Taking into account the number of trials,
and the other significant peaks neighboring this period,
the chance-probability of detecting such a modulation is estimated
to be $<$0.5\%. Given the low strength of the modulation it is impossible
to estimate the uncertainty on the period by fitting the pulse arrival
times. We instead estimate the uncertainty from the width of the
peak to be $\sim$$2\times 10^{-6}$~s. 
The lightcurve folded over the best-period was fit with a
sine curve to give a semi-amplitude of $2.1\pm0.8$\%.

\begin{figure}
\begin{center}
\hbox{\includegraphics[width=5.5cm,angle=-90]{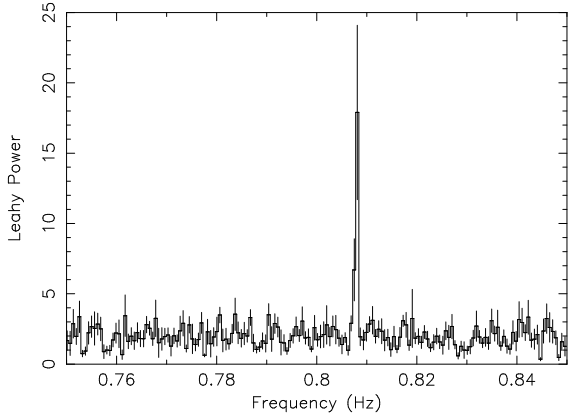}}
\caption[]{The power spectrum of \src\ obtained with the RXTE PCA 
during the anomalous low-state showing the clear detection of
pulsations}  
\label{fig:fft}
\end{center}
\end{figure}

To confirm this result public RXTE
Proportional Counter Array (PCA)
data of \src\ obtained on 1999 April 26 09:42 to 15:14~UTC
(close to an expected main-on) were analyzed. The exposure
time is 10.5~ks. A 2--60~keV lightcurve
was extracted from the event data with a resolution of 0.05~s.
Barycentering was performed with the {\sc fxbary} utility and the 
conversion to the system center of mass was performed 
as above. A clear modulation is detected at $>$99.999\% confidence
with a period of 1.237754(6)~s, confirming the
BeppoSAX discovery (see Fig.~\ref{fig:fft}). 
The (1$\sigma$) uncertainty was estimated by
fitting the arrival times of 8 sets of averaged pulses.

\subsection{X-ray Spectrum}
\label{subsec:spec}

Two sets of NFI spectra were extracted corresponding to the
non-eclipsed and eclipsed intervals in Fig.~\ref{fig:lc}. 
The summing of all the non-eclipse data is justified by the lack of
change in hardness ratio.
The spectra were rebinned to oversample the full
width half maximum (FWHM) of the energy resolution by
a factor 3 and to have additionally a minimum of 20 counts 
per bin to allow use of the $\chi^2$ statistic. 
Data were selected in the energy ranges
0.1--4.0~keV (LECS), 1.8--10~keV (MECS), 8.0--20~keV (HPGSPC),
and 15--40~keV (PDS).
Factors were included in the spectral fitting to allow for normalization 
uncertainties between the instruments. These were constrained
to be within their usual ranges during fitting. 

\begin{table}
\begin{center}
\caption[]{Partial covering fit to the non-eclipse
NFI spectrum. The energies of the narrow Fe-L lines were
fixed at values in Mihara \& Soong (\cite{m:94}).
f is the fraction of flux that undergoes extra absorption
${\rm N_{PCF}}$. A distance of 6.6~kpc is assumed}
\begin{tabular}{ll}
\hline
\noalign {\smallskip}
Parameter              &  \\
\hline
\noalign {\smallskip}
$\alpha$               & $0.63 \pm 0.02$ \\
${\rm E_c}$ (keV)      & $18 \pm 0.4$ \\
${\rm E_{fold}}$ (keV) & $9.0 \pm 0.8$ \\
Blackbody kT (keV)     & $0.068 \pm 0.015$ \\
Equiv. BB radius (km)  & $96 \pm 7 $ \\
\nh\ ($\times 10^{19}$ atom cm$^{-2}$) & $13.6 \pm ^{2.9} _{6.1}$ \\ 
f                      & $0.28 \pm 0.03 $ \\
${\rm N_{PCF}}$ ($\times 10^{22}$ atom cm$^{-2}$) & $27  \pm 7 $\\
Fe-L 0.91 keV EW (eV)  & $44 \pm ^{47} _{40}$ \\
Fe-L 1.06 keV EW (eV)  & $49 \pm 41 $ \\
Fe-K line energy (keV) & $6.57 \pm 0.04$ \\ 
Fe-K line FWHM (keV)   & $0.60 \pm 0.11 $ \\
Fe-K line EW (keV)     & $0.665 \pm 0.07$ \\
Intensity (0.1--10~keV; erg cm$^{-2}$ s$^{-1}$) & $4.5\times 10^{-11}$ \\
$\chi ^2$/dof          & 160.5/123 \\
\noalign {\smallskip}                       
\hline
\label{tab:spectral_fits}
\end{tabular}
\end{center}
\end{table}

The 0.1--10~keV spectrum of the \src\ main-on state and parts of the
short-on state are well 
described by a model consisting of an absorbed
power-law and blackbody continuum
together with two narrow lines with energies fixed at the
best-fit energies of Mihara \& Soong (\cite{m:94}) of 
0.91~keV and 1.06~keV and a broad line at $\sim$6.5~keV
(Oosterbroek et al. \cite{o:99}; hereafter O99). At higher energies
a cutoff is evident, together with cyclotron absorption
which may be modeled by a Gaussian feature in absorption
(e.g., Dal Fiume et al. \cite{df:98}). Above an energy ${\rm E_c}$
the cutoff is modeled as ${\rm \exp((E_c - E)/E_{fold})}$,
where E is energy in keV. The ``standard model'' used
in O99 was therefore modified with these additional high-energy
features and fit to the non-eclipse NFI spectrum. This did not give
a satisfactory result with a $\chi ^2$ of 335 for 142 dof.
In addition, the best-fit power-law photon index
$\alpha$ is 0.2, much harder than expected. 

This spectral shape is similar to that 
obtained towards the end of the short-on state, or in the low-state
by O99 where the effects of significant absorption are clearly seen 
as a change in spectral shape and increased curvature in the 
1--3 keV energy range implying absorption
of $\approxgt$$10^{22}$~atom~cm$^{-2}$. 
However, substantial flux remained $\approxlt$0.5 keV, which should be 
{\it completely} absorbed with such a high absorption. 
This suggests the presence of separate ``scattering'' 
and ``absorbing'' regions which O99 model as partial covering
using the {\sc pcfabs} model in {\sc xspec}.
Here a fraction, f, of the emission undergoes extra absorption,
${\rm N_{PCF}}$, while the rest is absorbed by a low value of
\nh, as before. 
This partial covering model, modified as above at high-energies,
gives a much better fit to the non-eclipse 
spectrum with a $\chi^2$ of 160.5 for 123 dof 
(Table~\ref{tab:spectral_fits}). 
The best-fit value
of $\alpha$ is now $0.63 \pm 0.02$, which is still flatter than the
main on-state spectrum where $\alpha \approx 0.9$.
The equivalent blackbody radius 
is $96 \pm 7$~km and the ratio of 0.1--10.0~keV flux in the blackbody
compared to the power-law is 11\%. 

Due to the lower count rate and integration time (the MECS exposure is
only 6.9~ks) the quality of the eclipse spectrum is much worse than
that of the non-eclipse interval. The NFI spectrum can be fit by a
simple power-law model to give $\alpha = 0.77 \pm 0.22$ and
a $\chi ^2$ of 24.6 for 19 dof.  If a blackbody with its temperature
fixed at 0.07~keV is included then the $\chi ^2$ reduces
to 13.5 for 18 dof implying that the blackbody is still
present during the eclipse.  The best-fit value of $\alpha$ is
$0.76 \pm 0.25$ and the equivalent blackbody radius is $\sim$50~km.

\begin{figure}
\begin{center}
\hbox{\includegraphics[width=9.4cm,angle=0]{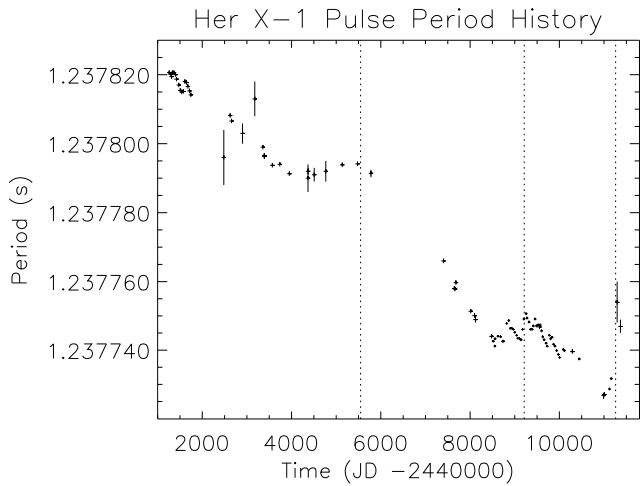}}
\caption[]{The pulse period history of \src\ from
Bildsten et al. (\cite{b:97}) updated using publically
available BATSE results. The pulse periods determined here are the
last two points. The approximate times of the 3 known
anomalous low-states of \src\ are indicated with dotted lines}
\label{fig:history}
\end{center}
\end{figure}

\section{Discussion}

We report on a BeppoSAX observation of \src\ during an
anomalous low-state. The 0.1--40~keV lightcurves show a gradual
reduction in count rate before the eclipse and then a more rapid
rise following egress. There is significant flux remaining during
eclipse. The shapes of the BeppoSAX lightcurves are similar to those
reported by Parmar et al. (\cite{p:85}) during a much shorter (8~hr) EXOSAT
observation centered on an eclipse during an earlier anomalous
low-state, although it is possible that the asymmetry in the lightcurves
is reversed. Parmar et al. (\cite{p:85}) also measure a very hard
2--10~keV spectrum with $\alpha = 0.52$ and a $6.29 \pm 0.25$~keV Fe line 
with an EW of $1.2 \pm 0.4$~keV. The
blackbody component was also detected by the low-energy imaging
telescopes on EXOSAT. 

The non-eclipse 0.1--40~keV \src\ spectrum
can be well fit with a blackbody and cutoff power-law
continuum model, together with Fe-L and Fe-K emission features,
a cyclotron absorption feature, and partial covering.
Partial covering of the \src\ spectrum is also required
during the late phases of the short-on state (O99), during the
1993 anomalous low-state (Vrtilek et al. \cite{v:94}) and during the
low-state (Mihara et al. \cite{m:91}).
The best-fit low-state Fe-K line energy, FWHM
and EW of $6.53 \pm 0.08$~keV, $0.8 \pm 0.4$~keV, and $1.0 \pm
0.1$~keV, respectively (Mihara et al. \cite{m:91})
are all remarkably similar to those reported
here (see Table~\ref{tab:spectral_fits}), whereas Vrtilek et al. (\cite{v:94})
report a weak line with an energy and EW of $6.31 \pm 0.1$~keV and 
$90 \pm 30$~eV
during the 1993 anomalous low-state. Thus, the anomalous low-state
spectrum observed here is indistinguishable from that 
observed during the more commonly occurring low-state.

The pulsations discovered during the anomalous low-state have a
semi-amplitude of $2.1\pm0.8$\%, consistent with the previous upper
limits obtained by Mihara \& Soong (\cite{m:94}) during the 1993 
anomalous low-state ($<$1.5\%) and Mihara et al.\ (\cite{m:91}) 
during the low-state ($<$2.4\%).
The pulse periods measured here indicate that \src\ underwent an interval
of rapid spin-down at around the time of the start of the anomalous low-state
(Fig.~\ref{fig:history}). A similar occurrence is evident in the case of the
1993 anomalous low-state (see Vrtilek et al. \cite{v:94}), while
the 1983 anomalous low-state appears to occur close to the end of
an extended interval of spin-down. These measurements therefore strengthen
the link between intervals of spin-down and the occurrence of anomalous
low-states.

Several pulsars exhibit transitions between intervals of spin-down and 
spin-up (e.g., Bildsten et al. \cite{bi:97}). Van Kerkwijk et al.
(\cite{vk:98}) 
propose that this behavior results from the accretion disk becoming
warped to the extent that the inner region becomes tilted by 
$>$${\rm 90^\circ}$, resulting in a retrograde flow and a spin-down torque.
This model implies that when spinning-down the X-ray source
would be mostly observed {\it through} the accretion disk. This is
unlikely to produce a simple increase in absorption since the inner
disk regions are likely to be substantially ionized and scattering may
play an important role. Support for this model comes from 
observations of the 7.7~s pulsar 4U\thinspace1626-67. During spin-up the 
power-law spectrum has $\alpha = 1.5$ while during 
spin-down $\alpha = 0.4$ (van Kerkwijk et al. \cite{vk:98}).
The spin-down spectrum can also be fit using a partial
covering model with 0.86 of the flux obscured 
by $6 \times 10^{23}$~atom~cm$^{-2}$. Thus, the 
4U\thinspace1626-67 spectral hardening 
and torque reversals are remarkably similar to those observed 
here from \src\ suggesting a common mechanism.
This implies that anomalous low-states
may result from an unusually large amount of warp in the \src\ accretion
disk. Since the accretion disk partially shadows HZ~Her from the X-ray
source, this change may produce a noticeable effect in the optical and
UV spectra of HZ~Her.

\acknowledgements 
The BeppoSAX satellite is a joint Italian and Dutch programme.

{}

\begin{thebibliography}{}

\bibitem[1997]{bi:97} Bildsten L., Chakrabarty D., Chiu J., et al., 1997, 
ApJS 113, 367

\bibitem[1997]{b:97} Boella G., Butler R.C., Perola G.C., 
et al., 1997, A\&AS 122, 299

\bibitem[1973]{b:73} Boynton P.E., Canterna R., Crosa L., Deeter J.,
Gerend D., 1973, ApJ 186, 617 

\bibitem[1998]{df:98} Dal Fiume D.,  Orlandini M., Cusumano G., et al., 1998
A\&A 329 L41

\bibitem[1991]{d:91} Deeter J.E., Boynton P.E., Miyamoto S., et al., 1991, 
ApJ 383, 324

\bibitem[1983]{d:83} Delgado A.J., Schmidt H.U., Thomas H.C., 1983,
A\&A 127, L15

\bibitem[1976]{g:76} Gerend D., Boynton P.E., 1976, ApJ 209, 562

\bibitem[1976]{j:76} Jones C.A., Forman W., 1976, ApJ 209, L131

\bibitem[1999]{l:99} Levine A.M., Corbet R., 1999, IAU Circ. 7139

\bibitem[1999]{m:99} Margon B., Deutsch E.W.,  Leinhardt Z.M., 
Anderson S.F., 1999, IAU Circ. 7144

\bibitem[1991]{m:91} Mihara T., Ohasi T., Makishima K., et al., 1991,
PASJ 43, 501

\bibitem[1994]{m:94} Mihara T., Soong Y., 1994, In: Makino F. (ed.) Proc.
of New Horizon of X-ray astronomy. Universal Academy Press, Tokyo, p.~419

\bibitem[1999]{o:99} Oosterbroek T., Parmar A.N., Dal Fiume D., et al.,
1999, A\&A submitted (O99)

\bibitem[1985]{p:85} Parmar A.N., Pietsch W., McKechnie S., et al., 
1985, Nat 313, 119

\bibitem[1999]{s:99} Scott D.M., Leahy D.A., 1999, ApJ 510, 974

\bibitem[1998]{vk:98} Van Kerkwijk M.H., Chakrabarti D., Pringle J.E.,
Wijers R.A.M.J., 1998, ApJ 499, L27

\bibitem[1994]{v:94} Vrtilek S.D., Mihara T., Primini F.A., et al.,
1994, ApJ 436, L9

\end{thebibliography}
\end{document}